\documentclass[prb,twocolumn,showpacs,superscriptaddress]{revtex4}

\usepackage{amsmath,amssymb,bm}
\usepackage{graphicx}
\usepackage{color}
\usepackage[colorlinks,linkcolor=blue,anchorcolor=blue,citecolor=blue,urlcolor=blue]{hyperref}

\renewcommand{\v}[1]{{\bf #1}}

\def\eqa{\begin{eqnarray}}
\def\eea{\end{eqnarray}}
\newcommand{\eq}{\begin{equation}}
\newcommand{\ee}{\end{equation}}
\newcommand{\nn}{\nonumber\\}
\newcommand{\Eq}[1]{Eq.~(\ref{#1})}
\newcommand{\<}{\langle}
\renewcommand{\>}{\rangle}

\renewcommand{\Im}{{\rm Im}}
\renewcommand{\Re}{{\rm Re}}
\newcommand{\p}{\partial}

\newcommand{\ua}{\uparrow}
\newcommand{\da}{\downarrow}
\newcommand{\ra}{\rightarrow}

\newcommand{\al}{\alpha}

\newcommand{\del}{\delta}
\newcommand{\Del}{\Delta}
\newcommand{\eps}{\epsilon}
\newcommand{\veps}{\varepsilon}

\newcommand{\La}{\Lambda}

\newcommand{\si}{\sigma}

\newcommand{\vphi}{\varphi}

\newcommand{\cO}{ {\cal O} }
\newcommand{\cP}{ {\cal P} }

\begin{document}

\title{Superconductivity, pair density wave, and Neel order in cuprates}

\author{Li-Han Chen}
\affiliation{National Laboratory of Solid State Microstructures \& School of Physics, Nanjing
	University, Nanjing, 210093, China}

\author{Da Wang}
%\email{dawang@nju.edu.cn}
\affiliation{National Laboratory of Solid State Microstructures \& School of Physics, Nanjing
	University, Nanjing, 210093, China}
\affiliation{Collaborative Innovation Center of Advanced Microstructures, Nanjing University, Nanjing 210093, China}

\author{Yi Zhou}
\affiliation{Beijing National Laboratory for Condensed Matter Physics \& Institute of Physics, Chinese Academy of Sciences, Beijing 100190, China}
\affiliation{CAS Center for Excellence in Topological Quantum Computation, University of Chinese Academy of Sciences, Beijing 100190, China}
\affiliation{Collaborative Innovation Center of Advanced Microstructures, Nanjing University, Nanjing 210093, China}

\author{Qiang-Hua Wang}
\email{qhwang@nju.edu.cn}
\affiliation{National Laboratory of Solid State Microstructures \& School of Physics, Nanjing
	University, Nanjing, 210093, China}
\affiliation{Collaborative Innovation Center of Advanced Microstructures, Nanjing University, Nanjing 210093, China}

%\date{\today}

\begin{abstract}
We investigate in underdoped cuprates possible coexistence of the superconducting order at zero momentum and pair density wave (PDW) at momentum ${\bf Q}=(\pi, \pi)$ in the presence of a Neel order. By symmetry, the $d$-wave uniform singlet pairing $dS_0$ can coexist with the $d$-wave triplet PDW $dT_{\bf Q}$, and the $p$-wave singlet PDW $pS_{\bf Q}$ can coexist with the $p$-wave uniform triplet $pT_0$. At half filling, we find the novel $pS_{\bf Q}+pT_0$ state is energetically more favorable than the $dS_0+dT_{\bf Q}$ state. At finite doping, however, the $dS_0+dT_{\bf Q}$ state is more favorable. In both types of states, the variational triplet parameters, $dT_{\bf Q}$ and $pT_0$, are of secondary significance. Our results point to a fully symmetric $\mathrm{Z_2}$ quantum spin liquid with spinon Fermi surface in proximity to the Neel order at zero doping, and to intertwined $d$-wave triplet PDW fluctuations and spin moment fluctuations along with the dominant $d$-wave singlet superconductivity at finite doping. The results are obtained by variational quantum Monte Carlo simulations.

\end{abstract}

%74.20.Rp  Pairing symmetries (other than s-wave)
%74.20.-z  Theories and models of superconducting state
%71.27.+a  Strongly correlated electron systems; heavy fermions

\maketitle

\section{introduction}
The investigation of the mechanism of high-temperature superconductivity in cuprates remains to be an exciting topic. One of the interesting proposals is Anderson's resonating valence bond (RVB) state. \cite{Anderson87-RVB}
A suitable Hamiltonian describing such a system is the one-band $t$-$J$ model. \cite{ZhangRice88-tJ} In this model, while the parent compound at half filling is automatically a Mott insulator for the charge degrees of freedom, the spin sector is much more intriguing. The RVB state may be viewed as a linear combination of configurations of the covering of spin singlets, a quantum spin liquid (QSL) with fractional spinon excitations. Chemical doping introduces mobile holes, leaving room for spin singlets to relocate and hence making the system a superconductor immediately.\cite{Lee06-review} Initially, an $s$-wave RVB state is proposed in view of the experimental robustness of superconductivity against impurity scattering,\cite{swave} but it is found later that the $d$-wave RVB state is energetically better,\cite{Gros88-VQMC,dwave}
and can actually be robust against impurities in doped Mott insulators because of charge renormalization. \cite{RMFT}

It should be pointed out, however, that in the undoped limit, there is a local charge-SU(2) symmetry following from the charge neutrality.\cite{su2} This symmetry relates various forms of RVB states. For example, the RVB state with a $\pi$-flux in the spinon hopping around a plaquette may be mapped to a state with $d$-wave pairing, etc. Such states are said to be gauge equivalent and describe the same spin liquid upon projection to the physical Hilbert space. The projective symmetry group (PSG) has been developed \cite{Wen02-PSG,Wen2004} to classify all possible and physically distinct spin liquids. The theory is built on the assumption that the spins are in a quantum disordered state. In a pure two-dimensional (2D) model, this might be a reasonable assumption since the continuous spin-SU(2) symmetry cannot be broken spontaneously at finite temperatures. However, one can still ask whether the moments could order at zero temperature, or in the ground state. Including inter-layer coupling can extend such a ground state order into finite temperatures. In fact, the Neel order is observed experimentally in the parent undoped compounds of cuprates. When the spin ordering is taken into account, the notion of spin liquid, with fractional spinon excitations, seems to be less well-defined, as spin-$\frac{1}{2}$ spinons would have been confined to form spin-1 magnons. However, inelastic neutron scattering experiments show that the spin excitations away from the Neel vector are broadened significantly, well beyond the linear spin-wave description. \cite{Piazza15-neutron} A recent interesting proposal is that even in the presence of Neel order, the spinons may be deconfined in a partial region of the Brillouin zone, \cite{pdeconf1,pdeconf2,Yu18-CPT} although the same phenomenon could also be understood in terms of magnon-magnon scattering.\cite{magnon1,magnon2,Yu18-CPT}  It is therefore interesting to consider spinon states on the background of the Neel order. 

\begin{figure}
	\includegraphics[width=0.8\columnwidth]{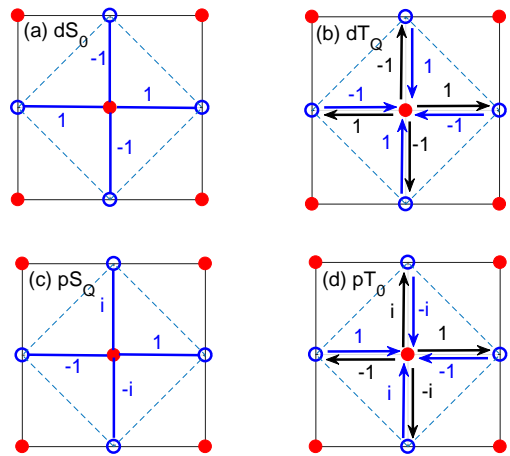}
	\caption{Illustration of the pairing function $\Del_{ij}$ on nearest-neighbor bonds, in the $d$-wave case (a) and (b), and the $p$-wave case (c) and (d), all on the background of a Neel order. The spin moments are opposite on the filled and open circles. The pairing pattern is shown only within a magnetic unit cell (dashed zone), but can be translated along the dashed lines to cover the entire lattice. For singlet pairing, $\Del_{ij}$ is symmetric under exchange of $i$ and $j$, hence its value is denoted on headless bonds in (a) and (c). For triplet pairing, the function is antisymmetric under the exchange of $i$ and $j$, hence its value is denoted by colored text on colored arrows in (b) and (d). An arrow starts at $i$ and points to $j$. From the transformation properties under rotations (about the sites) and translations (along the bonds), the pairing states are easily seen as: (a) $d$-wave singlet SC state $dS_0$, (b) $d$-wave triplet PDW state $dT_\v Q$, (c) $p$-wave singlet PDW state $pS_\v Q$, and (d) $p$-wave triplet SC state $pT_0$.}\label{fig:illustrate}
\end{figure}

The charge-SU(2) symmetry is broken at finite doping. The Neel order is weakened but persists at small finite doping. The effect of this order on superconductivity is also an intriguing topic.
In this case, the spin-SU(2) symmetry is broken down to O(2). As a result, there is no longer sharp distinction between spin-singlet and spin-triplet Cooper pairings, and there is room for coexistence, \cite{Lu14-topo,Gupta16-meanfield}
%\cite{Lu2014Underdoped,PhysRevLett.114.197001,2015arXiv150304969G}
although the relative weight is not dictated by symmetry. There is a residual point group $C_4$ (with respect to a site), which leaves the spin moment invariant in the Neel state. (In 3D, inversion is also a symmetry.) This symmetry dictates what kinds of singlet and triplet could coexist.
There are three irreducible representations for the $C_4$ group, namely, the 1D $A$ and $B$ representations, and the 2D $E$ representation. Since the completely symmetric $A$ representation is not a favorable pairing symmetry we will ignore it henceforth. The $B$ representation transforms as $d$-wave. The doubly degenerate $E$ representation transforms as $p$-wave. Therefore, the singlet and triplet should transform under $C_4$ identically either as $d$-wave or $p$-wave. These  possibilities are illustrated in Fig.\ref{fig:illustrate}. In the first case, the $d$-wave singlet Cooper pair at momentum $\v q=0$ (a), can coexist with a $d$-wave triplet at momentum $\v Q\equiv (\pi,\pi)$ (b). The latter is a pair density wave (PDW), namely, the Cooper pairing at the center-of-mass momentum of $\v Q$. In the second case, the $p+ip'$-wave singlet PDW at momentum $\v q=\v Q$ (c) can coexist with the same $p+ip'$-wave but triplet SC state (d). The four types of states are denoted in a self-explaining manner as $dS_0$, $dT_\v Q$, $pS_\v Q$ and $pT_0$. Interestingly, in the PDW states, an electron at momentum $\v k_1=\v k+\v Q$ pairs up with another one at $\v k_2=-\v k$. When $\v k$ is on the so-called umklapp surface (US), so will be both $\v k_1$ and $\v k_2$, related by mirror symmetry. 
%\cite{Lee14-Amperian, Zhang04-PDW, Wang15-PDW}
The scattering of such a Cooper pair across the US was argued to be the key mechanism that could generate not only the single-particle gap but also two-particle gap near the antinodes and hence the pseudogap. \cite{maurice} On the other hand, the chiral $pT_0$ state was recently proposed \cite{Lu14-topo} to be present on the background of the Neel order, in an effort to explain in the underdoped regime the opening of a mini gap in the quasiparticle excitations along the otherwise gapless (nodal) direction of the $dS_0$ state. \cite{Shen04-ARPES,Tanaka06-ARPES,Vishik12-ARPES,Razzoli13-ARPES,Peng13-ARPES}
%\cite{Peng2013Disappearance,Tanaka2006Distinct,Vishik2011Phase,PhysRevLett.110.047004,PhysRevB.62.4137,PhysRevB.69.054503,Gil2014Comprehensive}
Remarkably, if this were the case, the cuprate would be topologically nontrivial because of the chiral $p+ip'$ pairing. \cite{ReadGreen00-topo,QiZhang11-review}
%\cite{RevModPhys.83.1057,PhysRevB.38.931,PhysRevB.86.094512,Chang2013Protected,Alexei2009Periodic}

We are therefore motivated to investigate possible coexistence of superconducting (SC) order at zero momentum and PDW at momentum $\v Q$ in the presence of a Neel order. We use the variational quantum Monte Carlo (VQMC) to treat the strong correlation effects. Our main results are as follows. At half filling, we find the novel $pS_\v Q+pT_0$ state is energetically more favorable than the $dS_0+dT_\v Q$ state (including the conventional $dS_0$ state). We find that on its own the $pS_\v Q$ state is a fully symmetric $\mathrm{Z_2}$ QSL with spinon Fermi surfaces, but it is unstable toward Neel ordering. At finite low doping, however, the $dS_0+dT_\v Q$ state is more favorable. In both types of states, the variational triplet parameters, $dT_\v Q$ and $pT_0$, are of secondary significance since the ground state energy is degenerate with respect to varying levels of such variables (to some extent) upon optimization of the other parameters (including the chemical potential and hopping integrals). However, they do enhance the average Neel moment. Our results point to a novel QSL in proximity to the Neel order at zero doping, and to intertwined $d$-wave triplet PDW fluctuations and spin moment fluctuations along with the dominant $d$-wave singlet SC at finite doping.

In the rest of this paper, we specify the model and method in Sec.\ref{sec:mm}, discuss the results in Sec.\ref{sec:results}, and we draw conclusions and make perspective remarks in Sec.\ref{sec:summary}.

\section{model and method \label{sec:mm}}

We begin with the $t$-$J$ model on the square lattice, described by the Hamiltonian
\eqa H =&& -\sum_{n=1,2;\si}\sum_{\<ij\>\in N_n} t_{n}P(c_{i\si}^\dag c_{j\si}+{\rm h.c.})P \nn
    &&+ J\sum_{\<ij\>\in N_1} (\v S_i\cdot \v S_j - \frac{1}{4} Pn_i n_jP).\eea
Here $\si$ is the spin polarization, $N_{1,2}$ denote the first- and second-neighbor bonds with hopping integrals $t_1=t$ and $t_2=t'$, respectively; $c_{i\si}$ is the electron annihilation operator, $J$ is the Heisenberg spin exchange, $\v S_i$ is the local spin, and $n_i$ is the local density. The operator $P=\Pi_i (1-n_{i\ua}n_{i\da})$ projects away any double occupancy. As typical parameters for cuprates, we take $(t,t',J)=(0.4, -0.12, 0.13)$ eV. \cite{Lee06-review}
At half filling, the charge degrees of freedom are frozen, and the model reduces to the Heisenberg model. Upon hole doping, the doped holes may move without causing double occupancy, leading to metallicity and superconductivity.

The Hamiltonian includes infinitely strong correlations, due to the fact that no double-occupancy is allowed. In this work, we tackle the problem by variational quantum Monte Carlo (VQMC), which takes care of the no-double occupancy condition exactly. This method has been used extensively previously for the same system, \cite{Gros88-VQMC,Ogata99-VQMC,Ogata02-VQMC,Trivedi01-VQMC,Trivedi04-VQMC,Edegger07-review,Tan08-twomode}
yielding considerable insights into the Neel state at half filling and the uniform $d$-wave SC state at finite doping. More recently, the VQMC has been extended to deal with essentially unlimited number of variational parameters. \cite{Sorella05-newton,Umrigar05-newton,Morita}
Here we will go beyond the uniform $d$-wave ansatz yet limit ourselves to a handful of motivated parameters as we now describe. For later convenience, we introduce the Nambu basis $\psi_i^\dag = (c_{i\ua}^\dag, c_{i\da})$. The variational Hamiltonian can be written as,
\eqa H_v =&&\sum_{\<ij\>\in N_1} (\psi_i^\dag U_{ij}\psi_j + {\rm h.c.})-\sum_i \psi_i^\dag (\mu_z \eta_i+\mu\tau_3)\psi_i\nn && -\sum_{\<ij\>\in N_2} (\psi_i^\dag t_{2v}\tau_3 \psi_j+{\rm h.c.}).\label{eq:Hv}\eea
Henceforth $\tau_{1,2,3}$ are Pauli matrices in the Nambu (or particle-hole) basis. The second-neighbor hopping $t_{2v}$, the chemical potential $\mu$ and the exchange field $\mu_z$ are all variational parameters, and $\eta_i=\pm 1$ is a staggered sign for A/B sublattice. Furthermore, on the $N_1$ bonds,
\eqa U_{ij} = -t_{1v}\tau_3 + \Del_{ij}\tau^+ +\Del^*_{ji}\tau^-. \label{eq:uij}\eea Henceforth we fix $t_{1v}=1$ without loss of generality (since the only role of $H_v$ is to construct the trial wavefunction, see below), $\tau^{\pm} = (\tau_1 \pm  i\tau_2)/2$, and $\Del_{ij}$ is the real-space pairing function on a directed bond $\<ij\>$: For $\v b=\v r_j-\v r_i = (b_x, b_y)$,
\eqa \Del_{ij} = (b_x^2-b_y^2)(S_0 + T_\v Q \eta_i) + (b_x+ib_y) (S_\v Q \eta_i + T_0),\eea
where $S_0$ ($T_\v Q$) is the singlet (triplet) part of the $d$-wave pairing, and $S_\v Q$ ($T_0$) is the singlet (triplet) part of the chiral $p$-wave pairing. Note we assumed the triplets all have their $d$-vectors along $z$, the direction of the Neel moment. In this way, the total spin of a triplet Cooper pair is orthogonal to the Neel moment, a most favorable situation for triplets to develop on the Neel order induced by $\mu_z$. On the other hand, the momentum $\v Q=(\pi,\pi)$ in the PDW state is obvious from the staggered sign $\eta_i$ over the lattice. The four cases of the pairing function $\Del_{ij}$, when only one of the four coefficients in it is nonzero, are illustrated in Fig.\ref{fig:illustrate}. To summarize, we consider the set of variational parameters
\eqa &&x=\{\mu,\mu_z,t_{2v},S_0,T_\v Q\},\ \ \ d\mathrm{-wave\ case},\nn
     &&x=\{\mu,\mu_z,t_{2v},S_\v Q,T_0\}, \ \ \ p\mathrm{-wave\ case}.\eea
Since the $d$-wave and $p$-wave states belong to different irreducible representations of the $C_4$ group, we consider them separately. We assume all parameters in $x$ are real, as this turns out to gain energy better.

The normalized trial ground state is constructed as
\eqa |G\> = \frac{P|\psi_0\>}{\sqrt{\<\psi_0|P|\psi_0\>}}, \label{eq:G}\eea
where $|\psi_0\>$ is the ground state of the free variational Hamiltonian $H_v$ (which depends on the parameter set $x$), in the canonical ensemble with a definite total number of electrons, $N_e=N(1-\del)$, on the lattice. Here $N$ is the number of sites and $\del$ is the hole doping level. We use the standard Monte Carlo to calculate the average energy  density $E$ and the Neel order $m$,
\eqa E=\frac{1}{N}\<G|H|G\>,  \ \ m=\frac{1}{N}\sum_i \eta_i \<G| S_i^z|G\>.\eea
We optimize the parameter set $x$ automatically by adapting to our case the method proposed previously.\cite{Sorella05-newton,Umrigar05-newton,Morita}  More technical details can be found in the Appendix. We typically consider (tilted) square lattices with $N=82$, $128$, and $200$, and the finite size effect is found to be insignificant by casual check up to $N=800$. To stabilize the wavefunction, we apply antiperiodic boundary condition for fermions along one or both directions, and we add a tiny local singlet pairing that works even without using the antiperiodic boundary condition.

\section{Results and discussions \label{sec:results}}

\subsection{Variational results at zero doping}

\begin{figure}
	\includegraphics[width=\columnwidth]{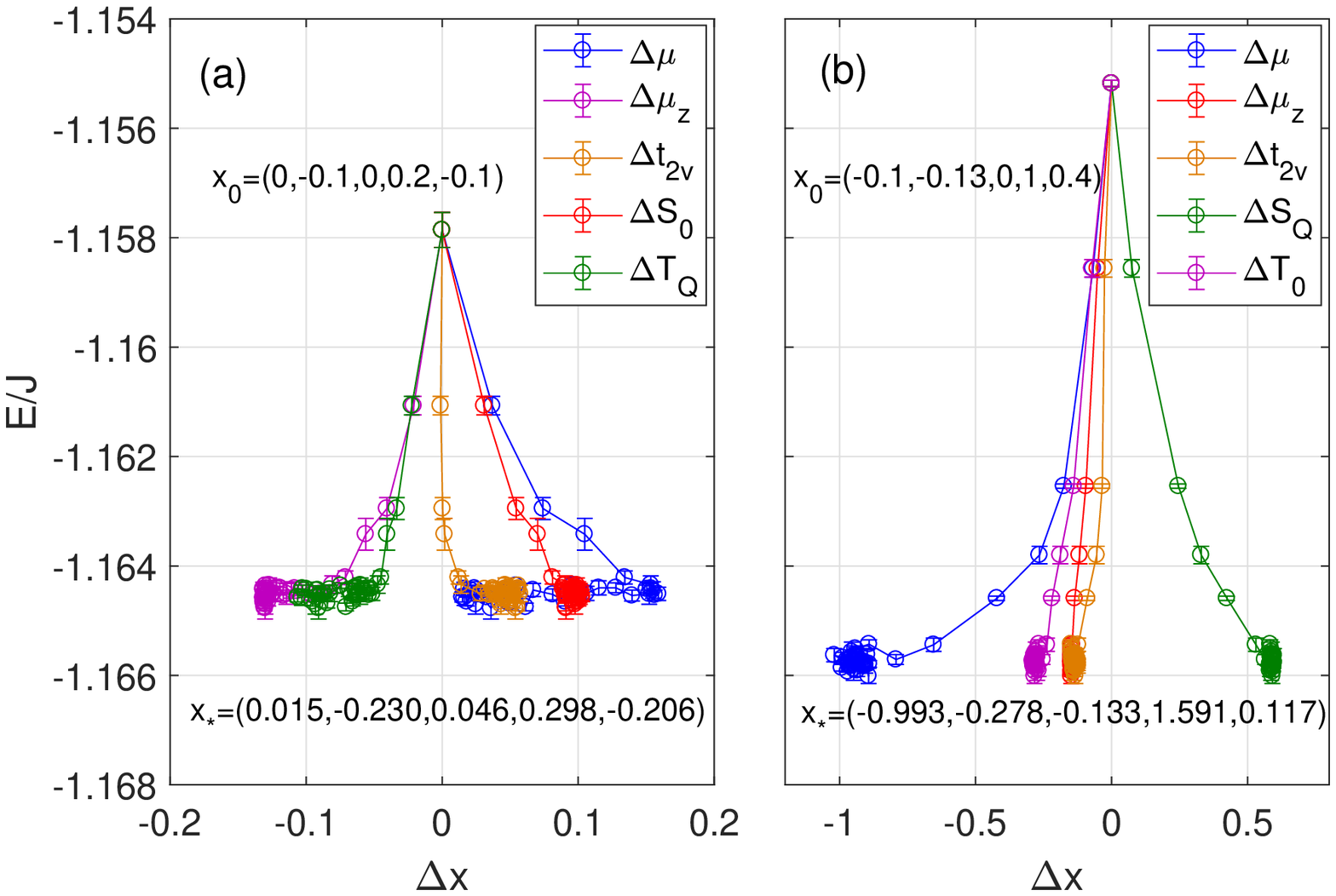}\vspace{0.4cm}
	\includegraphics[width=0.9\columnwidth]{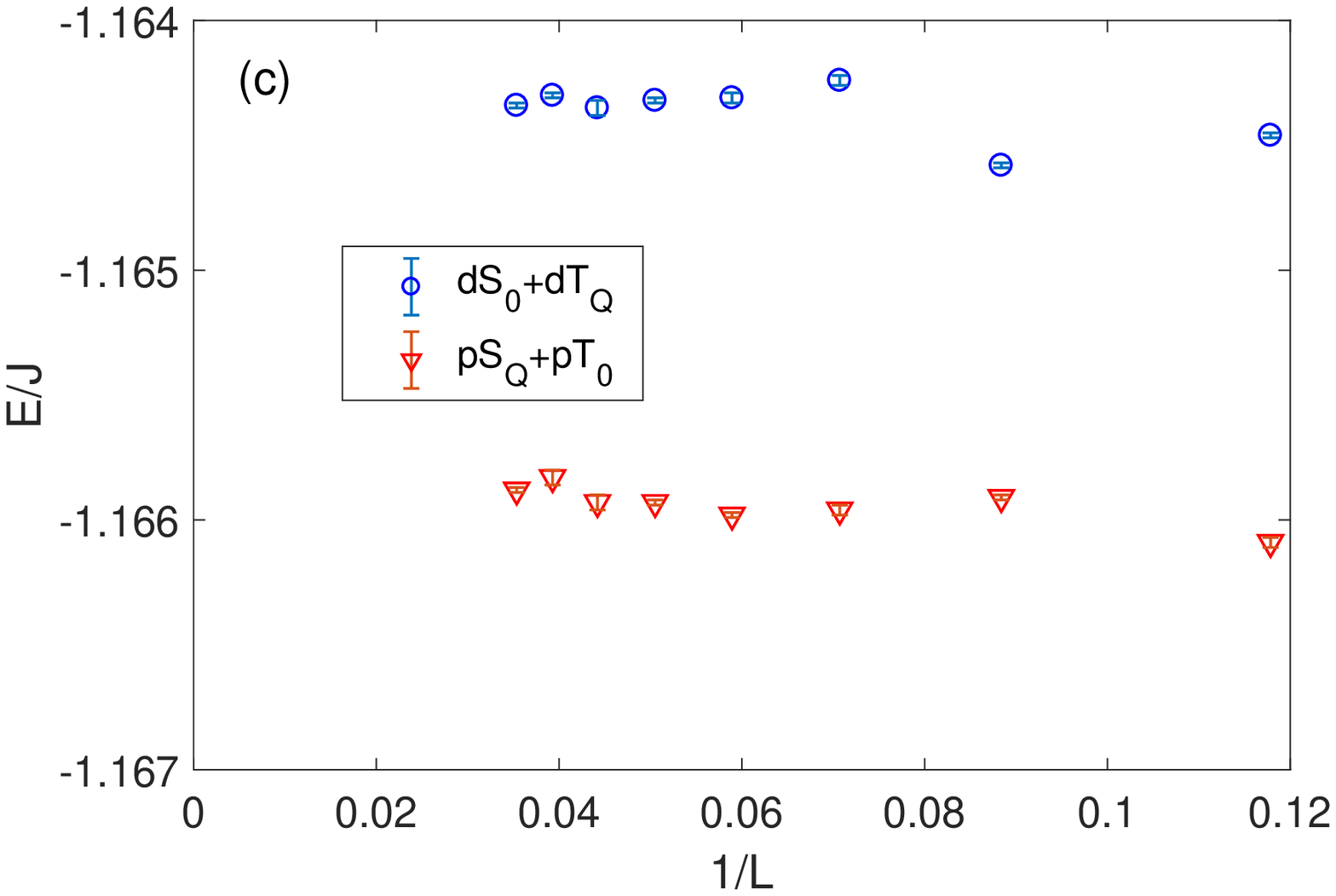}
	\caption{(a) and (b): Variation of the energy (per site) versus the change $\Del x$ of the parameter set $x$ during the automatic optimization for the undoped system for (a) the $d$-wave state with $x=(\mu,\mu_z,t_{2v},S_0,T_\v Q)$, and (b) the $p$-wave state with  $x=(\mu,\mu_z,t_{2v},S_\v Q,T_0)$. The process starts from an initial guess $x_0$, and reaches the stationary point $x_*$ with low energy very quickly. (c) Scaling of the optimized energy per site versus $1/L$, with $L=\sqrt{N}$ the linear lattice size, in the $dS_0+dT_\v Q$ state (open circles) and the $pS_\v Q+pT_0$ state (open triangles). In both cases the Neel order is included and optimized simultaneously.} \label{fig:undoped}
\end{figure}

In Fig.\ref{fig:undoped} we show the variation of the energy (per site) versus the change $\Del x$ of the parameter set $x$ during the automatic optimization of $x$ for the undoped system in (a) the $d$-wave state with $x=(\mu,\mu_z,t_{2v},S_0,T_\v Q)$, and (b) the $p$-wave state with  $x=(\mu,\mu_z,t_{2v},S_\v Q,T_0)$. The process starts from an initial guess $x_0$, and reaches quickly the stationary point $x_*$, where the energy is minimized. The lingering around $x_*$ provides a measure of the statistical error in energy and in the optimized parameters. In both cases, the parameter $\mu_z$ is finite, leading to Neel order, see Fig.\ref{fig:dope}(b) at zero doping. The best energy of the $d$-wave state in Fig.\ref{fig:undoped}(a) is about $E=-1.1645J$, or $\<\v S_i\cdot \v S_j\>=-0.3322J$ on $N_1$ bonds, consistent with that reported in the literature.\cite{YokoyamaShiba88-0.321,Gros88-VQMC,LeeFeng88-0.332,Lee06-review,Edegger07-review}
Interestingly, the optimized energy is even lower in the $p$-wave state in Fig.\ref{fig:undoped}(b), $E=-1.1660J$. Translated as $\<\v S_i\cdot \v S_j\>=-0.3330J$, the energy is so far the best using fermionic VQMC, and is fairly close to the results of bosonic VQMC ($-0.3344J$\cite{Liang-0.3344,Weng-0.3344}), Green's function QMC ($-0.3346J$ \cite{TrivediCepeley-0.3346}) and stochastic series expansion QMC ($-0.3347J$\cite{Sandvik97-QMC}).
Moreover, starting from a different initial guess $x_0$ we would obtain (not shown) a different optional $x_*$, but the energy is degenerate within statistical error. For example, in both cases of $d$-wave and $p$-wave, we obtain the same energy, respectively, by setting the triplet components to zero while optimizing the others. However, the energy is much poorer if we set the singlet components to zero instead. Specializing to the energetically more favorable $p$-wave case, we conclude the $p$-wave singlet PDW is relevant while the $p$-wave triplet is only marginal. This could be understood from the fact that locally the singlet on $N_1$ bonds gains energy from the anti-ferromagnetic spin exchange, while the triplet would be costly. The surprise is the singlet favors a center-of-mass momentum $\v Q$, in contrast to the usual uniform $d$-wave ansatz (related to our parameter $S_0$). The latter was widely assumed in previous fermionic VQMC.
To make sure that our result is not a finite-size artefact, we perform the optimization on lattices of various sizes, and the results are shown in Fig.\ref{fig:undoped}(c) as a scaling plot versus the linear size $L=\sqrt{N}$. The energy difference between the two types of states clearly survives in the limit of $1/L\ra 0$, suggesting insignificant finite-size effect. Therefore, our results point to a novel type of ground states with $p$-wave singlet PDW. An interesting question is whether such a pairing could persist at finite doping. We will come back to this point in the next section.

\subsection{A $\mathrm{Z_2}$ QSL with nested spinon Fermi surfaces and a pair of Dirac nodes}

At zero doping, we find that if we switch off the Neel order (by setting $\mu_z=0$), the optimized energy is $E=-1.1466J$ for the $pS_\v Q$ state, and $E=-1.1406J$ for the $dS_0$ state. The $pS_\v Q$ state is still better. The energy difference is far beyond statistical error. Furthermore, we can set $\mu=t_{2v}=0$ without affecting the optimized energy. In this case the variational Hamiltonian is only composed of the $U_{ij}$-terms in \Eq{eq:Hv}. We take a closer look into such a $pS_\v Q$ state to understand why it is better than $dS_0$, in terms of the PSG theory.\cite{Wen02-PSG,Wen2004}

There are staggered signs in $\Del_{ij}$ or in $U_{ij}$ in the $pS_\v Q$ state, see Fig.\ref{fig:illustrate}(c). This can actually be gauged away, given the exact charge neutrality (at zero doping) and hence local charge-SU(2) gauge invariance in VQMC. After the gauge transformation
\eqa U_{ij}\ra V_i U_{ij} V_j^\dag, \ \ \ V_i= e^{i\v Q\cdot \v r_i \tau_3/2},\eea
where $\v r_i=(x_i,y_i)$ is the coordinate vector, we obtain a uniform ansatz:
\eqa U_{i,i+\hat{x}}=-i\tau_0+S_\v Q \tau_2,\ \ U_{i,i+\hat{y}}=-i\tau_0+S_\v Q\tau_1,\label{eq:z2}\eea with $U_{ji}=U_{ij}^\dag$. The Wilson loop around an elementary plaquette counter-clockwise can be written as
\eqa W_{\cP(1234)}=\cP(U_1 U_2 U_3 U_4),\label{eq:loop}\eea where $\cP$ denotes a specific cyclic permutation for a given path, $U_{1,2}=i\tau_0+S_\v Q\tau_{1,2}$, and $U_{3,4}=U_{1,2}^\dag$. We find $W_{1234} = w e^{i\v F\cdot\tau/2}$, with $w$ a complex factor and $\v F\propto -(S_\v Q, -S_\v Q, 1)$. Since $[U_{1,2},W_{1234}]\neq 0$, the Wilson loops with respect to the same base point, obtainable by all cyclic permutations in \Eq{eq:loop}, carry noncolinear fluxoid, hence the invariant gauge group (IGG) is $\mathrm{Z_2}$. This IGG dictates that the distortion to the ansatz \Eq{eq:z2} in the form of $U_{ij}\ra U_{ij}e^{i\v a_{ij}\cdot\tau/2}$ is gapped for all directions of the gauge field $\v a_{ij}$. In comparison, there are massless U(1) gauge fluctuations in the $dS_0$ state.\cite{Wen02-PSG,Wen2004} Note that while the energy from VQMC is strictly invariant under local gauge transformation of $U_{ij}$, it does change under the gauge distortion if $\v a_{ij}$ cannot be gauged out. Therefore, the trial state $|G\>$ using a given ansatz $U_{ij}$ should work better when the gauge field is already massive, as in the case of \Eq{eq:z2}, which is gauge equivalent to the (nonmagnetic) $pS_\v Q$ state. 

\begin{figure}
	\includegraphics[width=0.9\columnwidth]{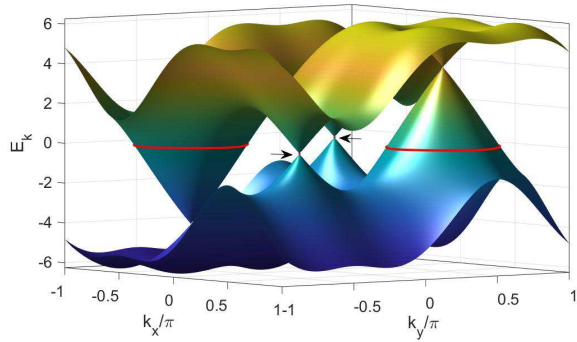}
	\caption{Spinon dispersion described by \Eq{eq:dispersion}, for $S_\v Q=1.7$ as an illustration. The thick red lines are the Fermi pockets and the arrows indicate a pair of Dirac nodes at the Fermi level.}\label{fig:fs}
\end{figure}

We further study the properties of our $\mathrm{Z_2}$ QSL. In the following we list the elements $g$ of the conventional symmetry groups, upon which $U_{ij}$ in \Eq{eq:z2} changes, but can be restored by a subsequent local gauge transformation $V_g$:
\eqa \begin{array}{ccccc} g= & P_x, & P_y, & P_{d}, & T. \\
	                     V_g= &   (-)^{x_i}i\tau_1, & (-)^{y_i}i\tau_2, & \frac{i}{\sqrt{2}}(\tau_1+\tau_2), & (-)^{x_i+y_i}.\end{array} \eea
In the above, $P_{x,y,d}$ are mirrors sending $x\ra -x$, $y\ra -y$ and $x\leftrightarrow y$, respectively, and $T$ is time reversal which acts on $U_{ij}$ as $i\tau_2 U_{ij}^* (-i\tau_2) = -U_{ij}$.\cite{Wen02-PSG,Wen2004} Note the four-fold rotation is identical to $P_xP_d$, so the point group $C_{4v}$ is automatically covered. We see that our spin liquid enjoys full physical symmetries. The set $\{V_g\cdot g\}$ forms the PSG of the spin liquid, and can be labeled as Z2A$xy(12)n$. \cite{Wen02-PSG,Wen2004} This PSG has not been realized in previous VQMC or self-consistent MF studies.

Using the ansatz in \Eq{eq:z2}, the variational Hamiltonian becomes
\eqa H_v=\psi_\v k^\dag[\veps_\v k\tau_0 + 2S_\v Q (\cos k_x \tau_2 + \cos k_y \tau_1)]\psi_\v k.\eea
Here $\veps_\v k=2(\sin k_x+\sin k_y)$. The quasiparticle dispersion is easily found to be
\eqa E_\v k = \veps_\v k\pm 2|S_\v Q|\sqrt{\cos^2 k_x + \cos^2 k_y}. \label{eq:dispersion}\eea
Clearly, the spinon excitation is gapless, and in fact there are two Fermi pockets enclosing $\pm \v Q/2$, see Fig.\ref{fig:fs}. In addition, at the Fermi level there are two Dirac nodes at $\pm (\pi/2, -\pi/2)$. Importantly, the spinon Fermi surfaces and Dirac nodes are protected by the above PSG, and this serves as an indicator of the quantum order in such a gapless $\mathrm{Z_2}$ QSL. \cite{Wen02-PSG,Wen2004} 

Finally, we discuss how the Neel order appears to lower the energy of the $pS_\v Q$ state further. Even if we assume the spinons are free from the coupling to the massive gauge fields, residual interactions between spinons can induce an instability in the presence of perfect nesting between the spinon Fermi surfaces. This results in a finite exchange field $\mu_z$ staggered over A- and B-sublattices. The quasiparticle dispersion becomes, in the folded Brillouin zone, $E_\v k'=\pm \sqrt{E_\v k^2+\mu_z^2}$. This eventually gaps out the spinons and breaks the spin rotation symmetry. We see an interesting example that the massiveness of gauge field fluctuations in a $\mathrm{Z_2}$ QSL is insufficient to guarantee its stability against magnetic ordering.

\subsection{Finite doping}

\begin{figure}
	\includegraphics[width=\columnwidth]{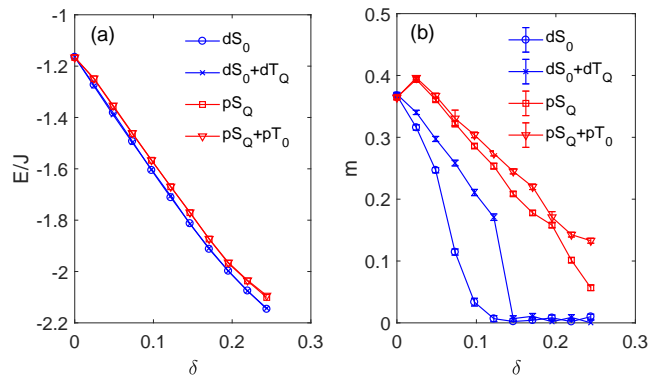}
	\caption{The doping dependence of (a) energy per site and (b) the Neel moment, in the various types of variational states. The results are obtained in a lattice with $N=82$. }\label{fig:dope}
\end{figure}

We have also performed systematic VQMC simulations at finite doping. In Fig.\ref{fig:dope}(a) we show the energy versus hole doping in the various variational states. We find the energy in the $dS_0$ and $dS_0+dT_\v  Q$ states are degenerate within statistical error, and so are the $pS_\v Q$ and $pS_\v Q+pT_0$ states. In contrast, the energy becomes poorer (not shown) if we get rid of the singlet components in both cases. This further enforces our view that the relevant pairing states are all singlets: $dS_0$ and $pS_\v Q$. Moreover, while in the undoped case we find the $pS_\v Q$ (or $pS_\v Q+pT_0$) state has lower energy, at finite doping, we find the $dS_0$ (or $dS_0+dT_\v Q$) state is systematically more favorable, down to the lowest nonzero doping we accessed. In fact, by linear interpolation the transition between these two types of states would be at a tiny hole doping level.
Fig.\ref{fig:dope}(b) shows the average Neel moment versus the hole doping. We find while the energy may be degenerate in the $d$-wave case, or in the $p$-wave case, as shown in (a), the moment differs if the triplet component is included. For example, the Neel moment is larger in the $dS_0+dT_\v Q$ state than the $dS_0$ state, and similarly for $pS_\v Q+pT_0$ versus $pS_\v Q$. 

Taking the lower-energy $dS_0+dT_\v Q$ state at finite doping, we believe the harmlessness of the triplet component $dT_\v Q$ to the optimized energy points to soft triplet PDW fluctuations and spin fluctuations in the underdoped regime. The dominant $dS_0$ is just the $d$-wave singlet pairing, well perceived in doped cuprates. The secondary $dT_\v Q$ state may become important at higher energy scales (e.g., above the superconducting transition temperature), where the umklapp scattering of such pairs may be the key process to generate the pseudogap near the antinodes in underdoped cuprates.\cite{maurice} Unfortunately, this is already beyond the scope of VQMC for the ground state.

On the other hand, our results indicate the $pS_\v Q+pT_0$ state is not favorable at finite doping. In particular, the chiral $pT_0$ state does not seem to help gain energy, at either zero or finite doping, with or without the Neel order. This appears to rule out the $pT_0$ state as a leading order parameter, which was proposed as an intriguing possibility on a tunable background of Neel order (and with tunable spin-exchange and Coulomb interaction on $N_1$-bonds).\cite{Lu14-topo}  Our VQMC treats the Neel order as one of the variational parameters for a given $t$-$J$ model. In another context, the $d$-wave pairing was shown by MFT to be stable, and the quasiparticle excitation becomes nodeless, on a strong Neel background.\cite{gm1,gm2} It is natural that the energetic ordering of the states in different symmetries depends on the starting model, and on the method by which the strong correlation effects are addressed.

\section{conclusion \label{sec:summary}}

We performed VQMC study of the $t$-$J$ model for cuprates, putting on equal footing the singlet and triplet pairings at zero as well as at momentum $\v Q=(\pi,\pi)$, along with the Neel moment. At zero doping we find the $p$-wave singlet PDW ($pS_\v Q$) to be favorable on the background of the Neel moment. We also discussed in terms of charge-SU(2) gauge theory that if the Neel order is ignored, the $pS_\v Q$ state is still favorable versus the $dS_0$ state, and describes a novel Z2A$xy(12)n$ QSL with spinon Fermi surfaces and Dirac nodes. However, this QSL is unstable toward Neel ordering. Above possibly a tiny doping level, we find the $dS_0+dT_\v Q$ state is established. We also find the uniform $p$-wave triplet ($pT_0$) is not a leading order at zero or finite doping. Our results point to a novel $\mathrm{Z_2}$ QSL in proximity to the Neel order at zero doping, and to intertwined $d$-wave triplet PDW fluctuations and spin moment fluctuations along with the dominant $d$-wave singlet SC at finite doping. The PDW state $dT_\v Q$ at finite doping may also provide the microscopic origin of the umklapp-scattering mechanism for the pseudogap \cite{maurice}. 

Two remarks are in order. First, we find our results at finite doping is qualitatively unchanged even if we set $t'=0$. Second, at finite doping, the magnetic order may become more complicated, such as the stripe order or even incommensurate magnetic order.\cite{Keimer15-review} Charge ordering is also possible. These orders reflect the fact that there are many intertwined or competing orders in strongly correlated systems such as the cuprates.\cite{Fradkin15-review} These orders require a larger set of variational parameters to accommodate, and are left for further studies.

\acknowledgments{The project was supported by the National Key Research and Development Program of China (under Grant Nos. 2016YFA0300401 and 2016YFA0300202), the National Natural Science Foundation of China (under Grant Nos.11574134, 11874205 and 11774306), and the Strategic Priority Research Program of Chinese Academy of Sciences (No. XDB28000000). The numerical simulations are performed in High-Performance Computing Center of Collaborative Innovation Center of Advanced Microstructures, Nanjing University.}

\section{Appendix}\label{appendix}

The ground state $|\psi_0\>$ of $H_v$ can be obtained straightforwardly as follows. Formally, we rewrite $H_v$ in the Nambu space,
\eqa H_v = \sum_{ij}\psi_i^\dag M_{ij} \psi_j,\eea
where $M_{ij}$ is a matrix in the Nambu as well as real spaces. Suppose $\vphi_n$ is the BdG eigenstate of $M$,
\eqa M\vphi_n =\eps_n \vphi_n,\eea
with eigen energy $\eps_n$. We write the eigenstate as $\vphi_n =(u_n,v_n)^t$, where $u_n$ is a column vector for the particle part, and $v_n$ a column vector for the hole part. There are in total $2N$ BdG states, and by particle-hole symmetry, the energies of these states appear pairwise in sign. We assume $\eps_{n}\leq 0$ for $n\leq N$, and construct the $N\times N$ matrices
\eqa U = (u_1, u_2, \cdots, u_N), \ \ V=(v_1, v_2, \cdots, v_N),\eea
and subsequently,
\eqa A=UV^{-1}.\eea
Then the ground state of $H_v$ can be written as
\eqa |\psi_0\>=\left(\sum_{ij} c_{i\ua}^\dag A_{ij} c_{j\da}^\dag\right)^{N_e/2}|0\>,\eea
where $N_e$ is the total number of electrons and $|0\>$ is the vacuum. The matrix $A$ depends on $x$ implicitly. In principle, $A$ can also be obtained first in the momentum space (with sublattice structure) and then transformed into the real space. But the above form is versatile. Finally, the normalized trial ground state for the $t$-$J$ model is given by \Eq{eq:G}.

We now describe how we minimize the energy $E$ by varying $x^\mu\in x$. Since we have a handful of variables to optimize, we try to perform the optimization automatically, rather than scanning over the parameter space {\em de forte}. The method has been developed in the literature. \cite{Umrigar05-newton,Sorella05-newton,Morita}
For self-completeness, here we review the method briefly, and we specify some subtleties to take care of in our case. Naively, the simplest way to update $x^\mu$ is the steepest descent method, $dx^\mu = - \p_\mu E dt$, where $\p_\mu E \equiv \p E/\p x^\mu$ and $dt$ is an artificial time step. However, this equation could cause instabilities in application. The reason is as follows. After $x$ has been updated as $x\to x+dx$, the wave function changes as $|G\> \ra |G\> + dx^\mu |\mu\>$, with $|\mu\> \equiv \p_\mu |G\>$. For brevity, summation over repeated indices is assumed henceforth, unless specified otherwise. Since the states $\{|\mu\>\}$ are not necessarily orthogonal, the effects of $\{dx^\mu\}$ are not independent, so that a small error in $dx^\mu$ may need a large $dx^\nu$ to compensate in a later stage. This is the root of instability. To make improvement, we form locally an orthonormal basis set $\{|a\>\}$ so that
\eqa dx^\mu|\mu\> = dx^a|a\>, \ \ \ dx^a = \<a|\mu\> dx^\mu \equiv e^a_\mu dx^\mu.\label{eq:dxa}\eea
We observe that
\eqa dx^\mu dx^\nu \<\mu|\nu\> \equiv dx^\mu dx^\nu g_{\mu\nu} = dx^{a*} dx^b \del_{ab}. \label{eq:metric}\eea
For real $dx^a$ and real metric tensor $g_{\mu\nu}=\<\mu|\nu\>$, this is exactly the invariant line element (squared) on a curved manifold. We will consider the simpler real case for a moment, because of its appealing geometrical interpretation, leaving the complex case afterwards. We now project the energy gradient into the orthogonal frame,
\eqa \p_a E = e_a^\mu \p_\mu E, \ \ \ e_a^\mu e_\nu^a =\del^{\mu}_{\nu}.\eea
We then apply the steepest descent method in the orthogonal frame,
\eqa dx^a = - \p_a E dt.\eea
We can now write this equation back to the $x^\mu$-frame,
\eqa e^a_\mu dx^\mu = -e^\mu_a\p_\mu E dt.\eea
Multiplying both sides by $e^a_\si$ and contracting $a$, we obtain
\eqa g_{\si\mu}dx^\mu = -\p_\si E dt,\ \ \ra\ \ dx^\mu = -g^{\mu\nu}\p_\nu E dt, \label{eq:update}\eea
where $g^{\mu\nu}$ is the inverse metric tensor. This form is explicitly covariant, and could have been obtained by an educated guess.

The metric tensor would be real if the variational parameters couple to operators diagonal in real space, such as the Jastrow factors for Hubbard models. This is unfortunately not our case. We therefore need to consider more general cases of $dx^a$ and $g_{\mu\nu}$. By definition we have
\eqa dx^{a*} = \<\mu|a\> dx^\mu.\label{eq:def}\eea
For a complex change $dx^a$, the steepest descent equation should be written as
\eqa dx^a = - \frac{\p E}{\p x^{a*}} dt.\eea
Going back to the $x^\mu$-frame, we obtain
\eqa \<a|\mu\> dx^\mu = -\frac{\p x^\mu}{\p x^{a*}} \p_\mu E dt.\eea
Multiplying both sides by $\<\nu|a\>$ and summing over $a$, we obtain
\eqa g_{\nu\mu}dx^\mu = -\frac{\p x^{a*}}{\p x^\nu}\frac{\p x^\mu}{\p x^{a*}} \p_\mu E dt = - \p_\nu E dt,\eea
where we used the relations, by definition,
\eqa \<\nu|a\>\<a|\mu\>=\<\nu|\mu\>=g_{\nu\mu},\ \ \<\nu|a\> = \frac{\p x^{a*}}{\p x^\nu}.\eea Therefore we are led to the same result as \Eq{eq:update}. A possible problem is the resulting $dx^\mu$ may be complex. A simple solution to the problem is to disregard the imaginary part of $g_{\mu\nu}$ without jeopardizing the invariant line element in \Eq{eq:metric}. The correctness and efficiency of doing so can only be judged by practice.

Moreover, even if the update is performed in the form of \Eq{eq:update}, instability could still appear if the metric tensor has small eigenvalues. For better stability in application, two further improvements are proposed in the literature. The first is to enlarge slightly the diagonal elements of the metric tensor, $g_{\mu\mu}\ra (1+\eps) g_{\mu\mu}$ (no summation over $\mu$), with a positive small number $\eps$. The second is to disregard the dangerous eigenmodes of the metric tensor. Since $g_{\mu\nu}$ is Hermitian and positive definite, we can expand its inverse as
\eqa g^{\mu\nu} \ra \sum_{k}^{'} \phi_k(\mu) \frac{1}{\La_k} \phi_k^*(\nu),\eea
where $\phi_k$ is an eigenvector of $g_{\mu\nu}$ with eigenvalue $\La_k$, and the primed summation excludes small eigenvalues that would blow up statistical errors in $\p_\mu E$ in \Eq{eq:update}.

We now specify how $\p_\mu E$ and $g_{\mu\nu}$ are calculated. We recall that $|G\>$ can be expressed as
\eqa |G\> =\frac{\sum_R \al_R |R\>}{\sqrt{\sum_R |\al_R|^2}}.\eea
Here $R=\{i_\ua,j_\da\}$ denotes a real-space configuration of the electrons without double-occupancy, and the coefficient $\al_R$ is the determinant of the sub-matrix $A_{i_\ua\in R, j_\da\in R}$. In the $R$-basis, we define the matrix $h_{RR'}=\<R|H|R'\>$, which is a sparse matrix. We take $|\al\>$ as the column vector composed of $\al_R$, or the wavefunction in the $R$-space, and define the average
$\<\cO\>=\<\al|\cO|\al\>/\<\al|\al\>$ for any (local or nonlocal) operator in $R$-space. We also define $l_\mu(R) = \p_\mu \ln \al_R$ formally as a local operator in the $R$-space. (This operator would be ill-defined if the phase of $\al_R$ does not vary smoothly with the parameter $x$. For example, an arbitrary global phase could be assigned to a wavefunction without any physical significance. To avoid such an ambiguity, in practice it is advisable to construct the wavefunction as a continuous fiber over the parameter space.) In these formal terms, we have
\eqa &&E = \frac{\sum_{RR'} \al_R^* h_{RR'}\al_{R'}}{\sum_R |\al_R|^2}\equiv \<h\>,\nn
&&\p_\mu E = \<h l_\mu\> - E \< l_\mu\> + {\rm c.c.}, \nn
&&g_{\mu\nu} = \<l_\mu^* l_\nu\> -\Re\<l_\mu\>\Re\<l_\nu\>\nn
&& \ \ \ \ \ \ \ +~i[\Im\<l_\mu\>\Re\<l_\nu\>-(\mu\leftrightarrow\nu)]. \eea
The last imaginary part drops out upon contraction with $dx^\mu dx^\nu$. The same is true in the imaginary part of $\<l_\mu^* l_\nu\>$. So eventually we simply take the real part of $g_{\mu\nu}$, as we argued previously. All of the above averages can be calculated conveniently by Monte Carlo. We checked that the automatic update works perfectly for the $dS_0+dT_\v Q$ state at any doping level, but meets instabilities for the $pS_\v Q+pT_0$ state at finite doping. In the latter case, the wavefunction seems to be too sensitive at some isolated points of $x$, resulting in jumps of the energy followed by steady decay. In the worst case we perform scanning over each direction of the parameter space to obtain reliable results.

%\bibliography{clh-tex}

\begin{references}

\bibitem{Anderson87-RVB} P. W. Anderson, Science {\bf 235}, 1196 (1987).

\bibitem{ZhangRice88-tJ} F. C. Zhang, and T. M. Rice, Phys. Rev. B {\bf 37}, 3759 (1988).

\bibitem{Lee06-review} P. A. Lee, N. Nagaosa, and X. G. Wen, Rev. Mod. Phys. {\bf 78}, 17 (2006).

\bibitem{swave} G. Baskaran, Z. Zou, and P. W. Anderson, Solid State Commun. {\bf 63}, 973 (1987).

\bibitem{Gros88-VQMC} C. Gros, Phys. Rev. B {\bf 38}, 931 (1988).

\bibitem{dwave} G. Kotliar and J. Liu, Phys. Rev. B {\bf 38}, 5124 (1988)

\bibitem{RMFT} Q. H. Wang, Z. D. Wang, Y. Chen, and F. C. Zhang, Phys. Rev. B {\bf 73}, 092507 (2006).

\bibitem{su2} I. Affleck, Z. Zou, T. Hsu, and P. W. Anderson, Phys. Rev. B {\bf 38}, 745 (1988).

\bibitem{Wen02-PSG} X. G. Wen, Phys. Rev. B {\bf 65}, 165113 (2002).

\bibitem{Wen2004}X. G. Wen, Quantum field theory of many body systems (Oxford University Press, New York, 2004).

\bibitem{Piazza15-neutron} B. D. Piazza, M. Mourigal, N. B. Christensen, G. J. Nilsen, P. Tregenna-Piggott, T. G. Perring, M. Enderle, D. F. McMorrow, D. A. Ivanov, and H. M. Ronnow, Nat. Phys. {\bf 11}, 62 (2015).

\bibitem{pdeconf1} N. S. Headings, S. M. Hayden, R. Coldea, and T. G. Perring, Phys. Rev. Lett. {\bf 105}, 247001 (2010).

\bibitem{pdeconf2} H. Shao, Y. Q. Qin, S. Capponi, S. Chesi, Z. Y. Meng, and A. W. Sandvik, Phys. Rev. X {\bf 7}, 041072 (2017).

\bibitem{Yu18-CPT} S. L. Yu, W. Wang, Z. Y. Dong, Z. J. Yao, and J. X. Li, Phys. Rev. B {\bf 98}, 134410 (2018).

\bibitem{magnon1} A. W. Sandvik and R. R. P. Singh, Phys. Rev. Lett. {\bf 86}, 528 (2001).

\bibitem{magnon2} M. Powalski, K. Schmidt, and G. Uhrig, SciPost Phys. {\bf 4}, 001 (2018).

\bibitem{Lu14-topo} Y. M. Lu, T. Xiang, and D. H. Lee, Nat. Phys. {\bf 10}, 634 (2014).

\bibitem{Gupta16-meanfield} A. Gputa, and D. Sa, Eur. Phys. J. B {\bf 89}, 24 (2016).

%\bibitem{Lee14-Amperian} P. A. Lee, Phys. Rev. X {\bf 4}, 031017 (2014).

%\bibitem{Zhang04-PDW} H. D. Chen, O. Vafek, A. Yazdani, and S. C. Zhang, Phys. Rev. Lett. {\bf 93}, 187002 (2004).

%\bibitem{Wang15-PDW} Y. Wang, D. F. Agterberg, and A. Chubukov, Phys. Rev. Lett. {\bf 114}, 197001 (2015).

\bibitem{maurice} Y. H. Liu, W. S. Wang, Q. H. Wang, F. C. Zhang, and T. M. Rice, Phys. Rev. B {\bf 96}, 014522 (2017).

\bibitem{Shen04-ARPES} K. M. Shen, T. Yoshida, D. H. Lu, F. Ronning, N. P. Armitage, W. S. Lee, X. J. Zhou, A. Damascelli, D. L. Feng, N. J. C. Ingle, H. Eisaki, Y. Kohsaka, H. Takagi, T. Kakeshita, S. Uchida, P. K. Mang, M. Greven, Y. Onose, Y. Taguchi, Y. Tokura, S. Komiya, Y. Ando, M. Azuma, M. Takano, A. Fujimori, and Z. X. Shen, Phys. Rev. B {\bf 69}, 054503 (2004).
% nodeless gap in LSCO, NCCO, Ca(2-x)Na(x)CuO2Cl2

\bibitem{Tanaka06-ARPES} K. Tanaka, W. S. Lee, D. H. Lu, A. Fujimori, T. Fujii, Risdiana, I. Terasaki, D. J. Scalapino, T. P. Devereaux, Z. Hussain, and Z. X. Shen, Science {\bf 314}, 1910 (2006).
% BSCCO

\bibitem{Vishik12-ARPES} I. M. Vishik, M. Hashimoto, R. H. He, W. S. Lee, F. Schmitt, D. Lu, R. G. Moore, C. Zhang,
W. Meevasana, T. Sasagawa, S. Uchida, K. Fujita, S. Ishida, M. Ishikado, Y. Yoshida, H. Eisaki, Z. Hussain, T. P. Devereaux, and Z. X. Shen, PNAS {\bf 109}, 18332 (2012).
% BSCCO

\bibitem{Peng13-ARPES} Y. Peng, J. Meng, D. Mou, J. He, L. Zhao, Y. Wu, G. Liu, X. Dong,
S. He, J. Zhang, X. Wang, Q. Peng, Z. Wang, S. Zhang, F. Yang, C. Chen, Z. Xu, T. K. Lee, and X. J. Zhou, Nat. Comm. {\bf 4}, 2459 (2013).
% BSLCO

\bibitem{Razzoli13-ARPES} E. Razzoli, G. Drachuck, A. Keren, M. Radovic, N. C. Plumb, J. Chang, Y. B. Huang, H. Ding, J. Mesot, and M. Shi, Phys. Rev. Lett. {\bf 110}, 047004 (2013).
% nodeless gap in LSCO

\bibitem{ReadGreen00-topo} N. Read, and D. Green, Phys. Rev. B {\bf 61}, 10267 (2000).

\bibitem{QiZhang11-review} X. L. Qi, and S. C. Zhang, Rev. Mod. Phys. {\bf 83}, 1057 (2011).

\bibitem{Ogata99-VQMC} A. Himeda, and M. Ogata, Phys. Rev. B {\bf 60}, R9935 (1999).

\bibitem{Ogata02-VQMC} A. Himeda, T. Kato, and M. Ogata, Phys. Rev. Lett. {\bf 88}, 117001 (2002).

\bibitem{Trivedi01-VQMC} A. Paramekanti, M. Randeria, and N. Trivedi, Phys. Rev. Lett. {\bf 87}, 217002 (2001).

\bibitem{Trivedi04-VQMC} A. Paramekanti, M. Randeria, and N. Trivedi, Phys. Rev. B {\bf 70}, 054504 (2004).

\bibitem{Tan08-twomode} F. Tan, and Q. H. Wang, Phys. Rev. Lett. 100, 117004 (2008).

\bibitem{Edegger07-review} B. Edegger, V. N. Muthukumar, and C. Gros, Adv. Phys. {\bf 56}, 927 (2007).

\bibitem{Umrigar05-newton} C. J. Umrigar, and C. Filippi, Phys. Rev. Lett. {\bf 94}, 150201 (2005).

\bibitem{Sorella05-newton} S. Sorella, Phys. Rev. B {\bf 71}, 241103 (2005).

\bibitem{Morita} S. Morita, R. Kaneko, and M. Imada, J. Phys. Soc. Jpn. {\bf 84}, 024720 (2015).

\bibitem{YokoyamaShiba88-0.321} H. Yokoyama, and H. Shiba, J. PHys. Soc. Jpn. {\bf 56}, 1490 (1987).

\bibitem{LeeFeng88-0.332} T. K. Lee, and S. P. Feng, Phys. Rev. B {\bf 38}, 11809 (1988).

\bibitem{Liang-0.3344} S. Liang, B. Doucot, and P. W. Anderson, Phys. Rev. Lett. {\bf 61}, 365 (1988).

\bibitem{Weng-0.3344} Z. Y. Weng, Y. Zhou, and V. N. Muthukumar, Phys. Rev. B {\bf 72}, 014503 (2005).

\bibitem{TrivediCepeley-0.3346} N. Trivedi, and D. M. Cepeley, Phys. Rev. B {\bf 40}, 2737(R) (1989).

\bibitem{Sandvik97-QMC} A. W. Sandvik, Phys. Rev. B {\bf 56}, 11678 (1997).

\bibitem{gm1} G. Y. Zhu and G. M. Zhang, EPL {\bf 117}, 67007 (2017). 

\bibitem{gm2} G. Y. Zhu, Z. Q. Wang, and G. M. Zhang, EPL {\bf 118}, 37004 (2017). 

\bibitem{Keimer15-review} B. Keimer, S. A. Kivelson, M. R. Norman, S. Uchida, and J. Zaanen, Nature {\bf 518}, 179 (2015).

\bibitem{Fradkin15-review} E. Fradkin, S. A. Kivelson, and J. M. Tranquada, Rev. Mod. Phys. {\bf 87}, 457 (2015).

\end{references}

\end{document}